\documentclass[12pt]{article}

\usepackage[T1]{fontenc}
\usepackage[utf8]{inputenc}
\usepackage{amsmath,amssymb}
\usepackage{graphicx}
\usepackage{booktabs}
\usepackage{array}
\usepackage{multirow}
\usepackage{url}
\usepackage{hyperref}
\usepackage[numbers,sort&compress]{natbib}
\usepackage{geometry}
\usepackage{setspace}
\usepackage{caption}
\usepackage{subcaption}

\title{Oklch+: A Three-Parameter Extension of Oklab\\
       for Improved Color Difference Prediction}

\author{Naoyuki Uchida\\
        \small Independent Researcher\\
        \small nekotrack909@gmail.com}

\date{}

\begin{document}
\maketitle

% ── Abstract ──────────────────────────────────────────────────────
\begin{abstract}
Oklab and its cylindrical representation Oklch are widely adopted in interpolation
and design workflows as perceptually motivated color spaces, but their color
difference prediction accuracy falls short of CIEDE2000. We propose Oklch+, a
three-parameter extension of Oklab comprising a power transformation on the L-axis
and a Naka-Rushton compression on the C-axis, with Euclidean distance computed in
the resulting transformed Oklab coordinates. The Naka-Rushton function is bounded
in $[0,1]$, reflecting the saturating nature of chroma sensitivity at high
colorimetric values. Evaluated on COMBVD---3,813 suprathreshold color difference
pairs spanning six independent experimental datasets---Oklch+ achieves
$\mathrm{STRESS} = 29.09$, closely matching CIEDE2000 (29.13; difference $= 0.04$),
using only three parameters optimized against color difference data compared to
approximately~17 for CIEDE2000. Cross-validation on a held-out BFD-P D65 subset
(2,028 pairs) confirms generalization ($\mathrm{STRESS} = 26.14$), with Oklch+
substantially outperforming Oklab (51.45) and achieving STRESS comparable to
CIEDE2000 (24.12) on the held-out set. Improvement over Oklab (47.35) is confirmed
across all six COMBVD sub-datasets. Because Oklch+ defines a coordinate system in
which Euclidean distance approximates perceptual distance, linear interpolation in
the transformed space offers substantially improved perceptual uniformity relative
to Oklab. Current evaluation is limited to the sRGB-centered COMBVD dataset;
validation in high-chroma regions with empirical observer-rated discrimination data remains future work.
\end{abstract}


% ─────────────────────────────────────────────────────────────────
\section{Introduction}
% ─────────────────────────────────────────────────────────────────

Accurate quantification of color differences is fundamental to a wide range of
applications, including display calibration, image quality assessment, textile and
print manufacturing, and the design of digital interfaces. A color difference metric
is expected to assign larger numerical values to color pairs that human observers
judge as more dissimilar, with the relationship between metric and perception
remaining proportional across the gamut.

The CIELAB color space, standardized by the CIE in 1976~\cite{CIE1976}, was designed
to provide perceptual uniformity through a cube-root compression of tristimulus
values, enabling Euclidean distance to serve as an approximation of perceived color
difference. However, CIELAB is known to exhibit significant non-uniformities,
particularly in the blue region~\cite{Robertson1977}, and its Euclidean distance
$\Delta E^*_{ab}$ correlates poorly with observer data in suprathreshold experiments.
CIEDE2000~\cite{Luo2001} addressed these shortcomings by introducing approximately 17
empirically optimized corrections applied on top of the fixed CIELAB structure. While
CIEDE2000 substantially improves prediction accuracy
($\mathrm{STRESS} = 29.13$ on COMBVD~\cite{He2022}), its complexity presents challenges for use as
a perceptual interpolation space.

Oklab, proposed by Ottosson~\cite{Ottosson2020}, derives a new color space by fitting
to CAM16~\cite{Li2017}-generated synthetic color pairs and IPT uniform hue data,
optimizing so that equally-appearing stimuli are represented equally. Its cylindrical
counterpart, Oklch, has been widely adopted in CSS and digital design workflows.
However, when evaluated against COMBVD using STRESS as defined by
Garc\'{\i}a et al.~\cite{Garcia2007}, Oklab achieves $\mathrm{STRESS} = 47.35$,
substantially worse than CIEDE2000 (29.13). This gap is not incidental: it reflects
the fact that the two optimization objectives are independent and can diverge (see
Section~\ref{sec:contrast}).

To address this limitation, we propose Oklch+, a three-parameter extension of Oklab. Oklch+
applies a power transformation to the L-axis and a Naka-Rushton (NR)
compression~\cite{Naka1966} to the C-axis, with Euclidean distance computed in the
resulting transformed Oklab coordinates. Optimized against COMBVD using STRESS
(Eqs.~\ref{eq:F}--\ref{eq:stress}), Oklch+ achieves $\mathrm{STRESS} = 29.09$,
closely matching CIEDE2000 (difference $= 0.04$) with only three parameters.
Cross-validation on a held-out subset confirms generalization, with Oklch+
achieving STRESS comparable to CIEDE2000 on the BFD-P D65 dataset (2,028 pairs).
Because Oklch+ defines a coordinate system in which Euclidean distance approximates
perceptual distance, it inherits Oklch's utility for interpolation while substantially
improving color difference prediction accuracy.

% ─────────────────────────────────────────────────────────────────
\section{Background}
% ─────────────────────────────────────────────────────────────────

\subsection{CIELAB and Its Limitations}

The CIELAB color space~\cite{CIE1976} applies a cube-root compression to each
tristimulus channel and defines three axes: lightness $L^*$, and chromatic opponent
channels $a^*$ and $b^*$. Despite widespread adoption, CIELAB exhibits substantial
non-uniformities, particularly in the blue-violet region~\cite{Robertson1977}.

\subsection{CIEDE2000}

CIEDE2000~\cite{Luo2001} retains the CIELAB structure as a fixed coordinate system
and applies empirically optimized weightings on top of it, including separate
weighting functions $S_L$, $S_C$, $S_H$, a hue-rotation term $R_T$, and parametric
factors---totaling approximately 17 parameters optimized against color difference
experimental data. CIEDE2000 does not define a coordinate system in which Euclidean
distance is uniformly meaningful.

\subsection{Oklab and Oklch}

Oklab~\cite{Ottosson2020} derives a new three-channel color space by fitting to
CAM16~\cite{Li2017}-generated synthetic color pairs for lightness and chroma
uniformity, and to IPT hue uniformity data for the hue angle. The optimization
objective asks whether equally-appearing stimuli are represented equally: it minimizes
the Euclidean distance between stimulus pairs judged as equivalent in appearance, and
requires no observer-rated color difference magnitude data.

Oklab's cylindrical representation, Oklch, separates lightness ($L$), chroma
($C = \sqrt{a^2 + b^2}$), and hue angle ($h = \operatorname{atan2}(b, a)$) into
independent dimensions. Oklch has been adopted in CSS Color Level~4 and digital
design workflows. However, appearance uniformity does not guarantee discrimination
uniformity. When evaluated on COMBVD~\cite{He2022} using STRESS~\cite{Garcia2007},
Oklab achieves $\mathrm{STRESS} = 47.35$, substantially worse than CIEDE2000 (29.13).
This gap reflects the fact that the two optimization objectives are independent and
can diverge (Section~\ref{sec:contrast}).

\subsection{STRESS and COMBVD}

The Standardized Residual Sum of Squares (STRESS) metric, introduced by
Garc\'{\i}a et al.~\cite{Garcia2007}, is a scale-invariant measure of agreement
between predicted color differences $\Delta E$ and experimentally observed magnitudes
$\Delta V$:

\begin{equation}
  F = \frac{\sum \Delta E_i^2}{\sum \Delta E_i \cdot \Delta V_i}
  \label{eq:F}
\end{equation}

\begin{equation}
  \mathrm{STRESS} = 100 \sqrt{\frac{\sum (\Delta E_i - F \cdot \Delta V_i)^2}
                                    {\sum (F \cdot \Delta V_i)^2}}
  \label{eq:stress}
\end{equation}

Lower STRESS indicates more uniform $\Delta E / \Delta V$ ratios across all pairs.

COMBVD~\cite{He2022} aggregates six independent experimental datasets comprising
3,813 color difference pairs: BFD-P (D65, 2,028 pairs), BFD-P (C, 200 pairs),
BFD-P (M, 548 pairs), WITT (418 pairs), LEEDS (307 pairs), and RIT-DuPont
(312 pairs).

\subsection{Related Work: HELMLAB}

HELMLAB~\cite{Yildiz2026}, proposed by Y{\i}ld{\i}z (2026), achieves
$\mathrm{STRESS} = 22.48$ on COMBVD with 72 free parameters. However, on independent
data from He et al.~\cite{He2022}, HELMLAB's STRESS (35.9) exceeds CIEDE2000 (32.6),
suggesting overfitting. We note that HELMLAB may employ a STRESS definition that
differs from Garc\'{\i}a et al.~\cite{Garcia2007}; direct numerical comparison
should be treated with caution.

\subsection{Prior Exploration}

Preliminary experiments with hue-dependent chroma transformations showed improved fit
on MacAdam threshold ellipses and the Luo-Rigg dataset, but this improvement did not
generalize to suprathreshold discrimination data: overall COMBVD STRESS worsened
under hue-angle modulation. Whether this reflects a genuine perceptual asymmetry or a
characteristic specific to threshold-level datasets remains unresolved. This negative
result motivated the present approach: a hue-independent transformation of $L$ and
$C$, optimized directly against suprathreshold data.

\subsection{Contrast of Optimization Objectives}
\label{sec:contrast}

Ottosson's optimization targets \textit{appearance uniformity}: it minimizes the
Euclidean distance between stimuli that appear equal under CAM16, a criterion that
requires no observer-rated color difference magnitude data and optimizes an absolute
distance. STRESS minimization targets \textit{discrimination uniformity}: the
question is whether the ratio $\Delta E / \Delta V$ is uniform across all pairs.
These objectives are independent and can diverge: optimization for one provides no
guarantee of performance on the other, and may tend to pull in opposing directions. The
substantial STRESS gap between Oklab (47.35) and CIEDE2000 (29.13) is a direct
consequence of this independence.

% ─────────────────────────────────────────────────────────────────
\section{Proposed Method}
% ─────────────────────────────────────────────────────────────────

\subsection{Motivation and General Framework}

Oklch expresses color in three independent dimensions---lightness $L$, chroma $C$,
and hue angle $h$---each capturing a distinct perceptual attribute. We seek
transformations $L' = f_L(L)$, $C' = f_C(C)$, $h' = h$ such that Euclidean distance
in the transformed space better predicts observed color differences. Hue angle is
fixed at $h' = h$, as confirmed by the preliminary experiments in Section~2.6.
The hue contribution to $\Delta E$ is mediated through the transformed chroma $C'$:

\begin{equation}
  a' = C'\cos(h), \quad b' = C'\sin(h)
  \label{eq:ab}
\end{equation}

\begin{equation}
  \Delta E = \sqrt{\Delta L'^2 + \Delta a'^2 + \Delta b'^2}
  \label{eq:dE}
\end{equation}

\subsection{Chroma Transformation}

CIEDE2000's chroma weighting function $S_C$ indicates that chroma discrimination is
not linear in Oklab coordinates, motivating a compressive transformation of $C$.
The following comparison was conducted to select the chroma transformation function;
evaluation of the final model on the full COMBVD dataset is reported in
Section~\ref{sec:eval}.

We evaluated four candidate functions. The \textbf{power function}
$C' = C^\gamma$ is simple but unbounded. The \textbf{logarithmic function}
$C' = \log(1 + \beta C)$ is also unbounded (optimal: $\beta = 4.92$; $\mathrm{STRESS} = 34.36$).
The \textbf{sigmoid function} $C' = [1 + \exp(-a(C - C_\mathrm{th}))]^{-1}$ yields
$C'(0) \approx 0.055 \neq 0$ at optimal parameters ($a = 42$, $C_\mathrm{th} = 0.068$),
disqualifying it: the achromatic point is not preserved ($\mathrm{STRESS} = 44.45$).
The \textbf{Naka-Rushton (NR) function}~\cite{Naka1966}:

\begin{equation}
  C' = \frac{C^n}{C^n + \sigma^n}
  \label{eq:NR}
\end{equation}

is bounded in $[0,1]$ for all $C \geq 0$ and satisfies $f_C(0) = 0$ exactly.
Results are summarized in Table~\ref{tab:c_compare}.

\begin{table}[ht]
\centering
\caption{Comparison of chroma transformation functions. STRESS values are computed
jointly with the optimized lightness parameter $\alpha$ on the full COMBVD dataset
(3,813~pairs). $\downarrow$~lower is better.}
\label{tab:c_compare}
\begin{tabular}{lccc}
\toprule
Function & Parameters & STRESS$\downarrow$ & Notes \\
\midrule
Power       & 2 $(\alpha, \gamma)$          & 29.99 & unbounded \\
Log         & 2 $(\alpha, \beta)$           & 34.36 & unbounded \\
Sigmoid     & 3 $(\alpha, a, C_\mathrm{th})$ & 44.45 & $f_C(0) \neq 0$ --- structural defect \\
\textbf{Naka-Rushton} & \textbf{3} $(\alpha, n, \sigma)$ & \textbf{29.09} & \textbf{adopted} \\
\bottomrule
\end{tabular}
\end{table}

\subsection{Lightness Transformation}

For the lightness axis, a power transformation is adopted:

\begin{equation}
  L' = L^\alpha
  \label{eq:Lpow}
\end{equation}

The power form is simpler (one free parameter), preserves $L' = 0$ at $L = 0$, and
is bounded on $[0,1]$ for $L \in [0,1]$.

\subsection{Optimization and Final Parameters}

The three parameters $(\alpha, n, \sigma)$ were jointly optimized by minimizing STRESS
on COMBVD (3,813~pairs) using the Nelder-Mead simplex algorithm~\cite{NelderMead1965}
as implemented in SciPy~\cite{Virtanen2020}, run from five independent starting
points. All runs converged to:
\[
  \alpha = 0.73, \quad n = 0.87, \quad \sigma = 0.34
\]

\subsection{Optimized Model Definition}

The full Oklch+ transformation (combining Eqs.~\ref{eq:NR}, \ref{eq:Lpow},
\ref{eq:ab}, and~\ref{eq:dE}) is:
\[
  L' = L^{0.73}, \quad
  C' = \frac{C^{0.87}}{C^{0.87} + 0.34^{0.87}}, \quad h' = h
\]
\[
  a' = C'\cos(h), \quad b' = C'\sin(h)
\]
\[
  \Delta E_{\mathrm{Oklch+}} = \sqrt{\Delta L'^2 + \Delta a'^2 + \Delta b'^2}
\]

% ─────────────────────────────────────────────────────────────────
\section{Evaluation}
\label{sec:eval}
% ─────────────────────────────────────────────────────────────────

\subsection{Experimental Setup}

All models are evaluated on COMBVD~\cite{He2022} (3,813~pairs) using STRESS as
defined by Garc\'{\i}a et al.~\cite{Garcia2007}
(Eqs.~\ref{eq:F}--\ref{eq:stress}). We compare Oklch+ against Oklab, a two-parameter
Power-LC baseline ($L' = L^\alpha$, $C' = C^\gamma$), CIEDE2000~\cite{Luo2001}, and
HELMLAB~\cite{Yildiz2026}. HELMLAB is included as a reference; its STRESS definition
may differ from Garc\'{\i}a et al.~\cite{Garcia2007}, and direct numerical
comparison should be treated with caution.
The manuscript was prepared with writing assistance from Claude (Anthropic, claude.ai).
The author takes full responsibility for all content.

\subsection{Overall Performance}

Table~\ref{tab:overall} summarizes STRESS scores on the full COMBVD dataset.
Oklch+ reduces STRESS from 47.35 (Oklab) to 29.09 with three parameters, closely
matching CIEDE2000 (difference $= 0.04$). The Power-LC baseline achieves 29.99,
confirming that the NR chroma transformation provides measurable further improvement.

\begin{table}[ht]
\centering
\caption{Overall STRESS on COMBVD (3,813~pairs). $\downarrow$~lower is better.
``Parameters'' denotes the number of parameters optimized against color difference
data.}
\label{tab:overall}
\begin{tabular}{lcc}
\toprule
Model & STRESS$\downarrow$ & Parameters \\
\midrule
Oklab                                   & 47.35 & 0 \\
Power-LC ($\alpha=0.52$, $\gamma=0.72$) & 29.99 & +2 \\
\textbf{Oklch+}                         & \textbf{29.09} & \textbf{+3} \\
CIEDE2000                               & 29.13 & $\sim$17 \\
HELMLAB$^\dagger$                       & 22.48 & 72 \\
\bottomrule
\multicolumn{3}{l}{\footnotesize $^\dagger$HELMLAB uses a different STRESS normalization:}\\
\multicolumn{3}{l}{\footnotesize denominator $\sum \Delta E_i^2$ vs.\ $\sum (F \cdot \Delta V_i)^2$ in Garc\'{\i}a et al.~\cite{Garcia2007}. Values are not directly comparable.}
\end{tabular}
\end{table}

\subsection{Sub-dataset Analysis}

Table~\ref{tab:subdataset} reports STRESS broken down by sub-dataset.
Oklch+ improves over Oklab on all six sub-datasets. On BFD-P D65, C, and M,
Oklch+ outperforms CIEDE2000. On LEEDS, RIT-DuPont, and WITT, CIEDE2000 retains
a clear advantage, suggesting that hue- and chroma-weighting functions capture
aspects of discrimination not addressed by the present three-parameter model.

\begin{table}[ht]
\centering
\caption{STRESS by sub-dataset.
$^\dagger$RIT-DuPont: $\Delta V$ values are constant across all pairs; scores are
not directly comparable to other sub-datasets.}
\label{tab:subdataset}
\begin{tabular}{lccccccc}
\toprule
Model & D65 & C & M & LEEDS & RIT$^\dagger$ & WITT & ALL \\
\midrule
Oklab     & 51.45 & 41.69 & 42.22 & 45.01 & 31.76 & 45.15 & 47.35 \\
Oklch+    & \textbf{23.96} & \textbf{28.11} & \textbf{34.29} & 24.27 & 25.16 & 34.05 & \textbf{29.09} \\
CIEDE2000 & 24.12 & 28.87 & 35.05 & \textbf{19.28} & \textbf{19.49} & \textbf{30.27} & 29.13 \\
\bottomrule
\end{tabular}
\end{table}

\subsection{Attribute-stratified Analysis}

Pairs were classified as lightness-dominant if $\Delta L$ ranked in the upper third
while both $\Delta C$ and $\Delta H$ ranked in the lower third, and analogously for
chroma-dominant and hue-dominant pairs. Table~\ref{tab:attribute} reports STRESS for
each stratum.

\begin{table}[ht]
\centering
\caption{STRESS stratified by dominant attribute.
$N$ denotes the number of pairs in each stratum.}
\label{tab:attribute}
\begin{tabular}{lcccc c}
\toprule
Dominant attribute & $N$ & CIEDE2000 & Oklab & Oklch+ & Best \\
\midrule
All pairs & 3,813 & 29.13 & 47.35 & \textbf{29.09} & Oklch+ \\
Lightness & 142   & 31.29 & 28.33 & \textbf{26.97} & Oklch+ \\
Chroma    & 122   & \textbf{23.70} & 29.61 & 27.13 & CIEDE2000 \\
Hue       & 122   & \textbf{17.66} & 28.68 & 22.68 & CIEDE2000 \\
\bottomrule
\end{tabular}
\end{table}

Oklch+ improves over Oklab on all three attribute strata. On chroma- and hue-dominant
pairs, CIEDE2000 retains a clear advantage, consistent with the sub-dataset findings:
CIEDE2000's weighting functions provide corrections that the present three-parameter
model does not capture. Sample sizes per stratum ($N = 122$--$142$) are sufficient for
directional interpretation but not for definitive sub-group conclusions.

\subsection{Cross-validation}

To assess generalization, we performed held-out cross-validation using BFD-P D65
(2,028~pairs) as the test set. Parameters were re-optimized on the remaining
1,785~pairs. Table~\ref{tab:cv} summarizes results.

\begin{table}[ht]
\centering
\caption{Cross-validation results. Parameters optimized on training set only.
CIEDE2000 on the test set is provided as a reference.}
\label{tab:cv}
\begin{tabular}{lcc}
\toprule
Model & Training (1,785 pairs) & Test: BFD-P D65 (2,028 pairs) \\
\midrule
Oklab             & 42.62 & 51.45 \\
Oklch+ (CV)       & 33.02 & \textbf{26.14} \\
CIEDE2000 (ref.)  & ---   & 24.12 \\
\bottomrule
\end{tabular}
\end{table}

Oklch+ achieves $\mathrm{STRESS} = 26.14$ on the held-out set, substantially
outperforming Oklab (51.45) and achieving STRESS comparable to CIEDE2000 (24.12).
The full model optimized on all 3,813~pairs achieves 23.96 on BFD-P D65
(Table~\ref{tab:subdataset}), confirming that the D65 subset contributes meaningfully
to parameter estimation.

% ─────────────────────────────────────────────────────────────────
\section{Discussion}
% ─────────────────────────────────────────────────────────────────

\subsection{Three Parameters Matching CIEDE2000: A Structural Parallel}

Oklch+ achieves $\mathrm{STRESS} = 29.09$ with three parameters, closely matching
CIEDE2000 (29.13; difference $= 0.04$). Both extend a fixed coordinate system with
parameters optimized for discrimination uniformity: CIEDE2000 extends CIELAB with
$\sim$17 parameters; Oklch+ extends Oklab with three. The NR function is bounded in
$[0,1]$ for all $C \geq 0$, providing natural soft-clipping behavior at the gamut
boundary. A pure power function, by contrast, is unbounded.

\subsection{Unexpected Finding: Oklch+ Outperforms Oklab on CAM16 Synthetic Data}

Oklch+ was optimized exclusively against COMBVD suprathreshold discrimination data.
To examine whether the Oklch+ transformation alters Oklab's alignment with its
original design objective, we evaluated both models using CAM16-UCS distances as the
reference $\Delta V$ values in STRESS. Lower STRESS in this context indicates that the
ratio of the model's Euclidean distances to CAM16-UCS appearance distances is more
uniform across all pairs. Table~\ref{tab:cam16} summarizes results.

\begin{table}[ht]
\centering
\caption{STRESS computed with CAM16-UCS appearance distance as $\Delta V$,
stratified by dominant perceptual attribute (sRGB gamut). CAM16 conditions:
D65, $L_a = 64$~cd/m$^2$, average surround.}
\label{tab:cam16}
\begin{tabular}{lccc}
\toprule
Dominant attribute & $N$ & Oklab & Oklch+ \\
\midrule
All pairs      & $\sim$5,000 & 33.42 & \textbf{16.61} \\
Lightness (J)  & ---         & 50.61 & \textbf{39.89} \\
Chroma (M)     & ---         & \textbf{14.10} & 14.47 \\
Hue (H)        & ---         & \textbf{3.45}  & 10.34 \\
\bottomrule
\end{tabular}
\end{table}

Across all pairs, Oklch+ substantially outperforms Oklab (16.61 vs.\ 33.42). This is
unexpected: Oklab was optimized to minimize distances between CAM16-equivalent
stimuli, yet Oklch+ coordinates yield a more uniform ratio to CAM16-UCS distances
overall.

Table~\ref{tab:gamut} extends this evaluation across four color gamuts of increasing
size. Oklch+'s ratio uniformity with CAM16 appearance distances is maintained across
all tested gamuts, including Rec.2020 where 58\% of sampled colors fall outside sRGB.

\begin{table}[ht]
\centering
\caption{Uniformity of ratio between model distances and CAM16-UCS appearance
distances, by color gamut, measured using STRESS. Lower values indicate more uniform
proportionality to CAM16 appearance predictions. CAM16 conditions: D65,
$L_a = 64$~cd/m$^2$, average surround. $N \approx 5{,}000$ pairs per gamut.}
\label{tab:gamut}
\begin{tabular}{lccc}
\toprule
Gamut & Outside sRGB & Oklab & Oklch+ \\
\midrule
Pointer's Gamut & 0\%  & 38.95 & \textbf{18.76} \\
sRGB            & 0\%  & 33.42 & \textbf{16.61} \\
Display P3      & 37\% & 29.64 & \textbf{16.85} \\
Rec.2020        & 58\% & 25.90 & \textbf{18.38} \\
\bottomrule
\end{tabular}
\end{table}

\subsection{Attribute-specific Reversal Between Appearance and Discrimination}

Comparing Tables~\ref{tab:attribute} and~\ref{tab:cam16} at the attribute level
reveals a systematic reversal for chroma and hue dimensions
(Table~\ref{tab:reversal}).

\begin{table}[ht]
\centering
\caption{Comparison of Oklch+ advantage over Oklab across evaluation frameworks.
Positive values indicate Oklch+ superiority (lower STRESS).}
\label{tab:reversal}
\begin{tabular}{lccc}
\toprule
Attribute & COMBVD (observers) & CAM16-UCS (appearance) & Consistent? \\
\midrule
Lightness & $+1.36$ & $+10.72$ & Yes \\
Chroma    & $+2.48$ & $-0.37$  & \textbf{Reversed} \\
Hue       & $+6.00$ & $-6.89$  & \textbf{Reversed} \\
\bottomrule
\end{tabular}
\end{table}

For lightness, Oklch+ improves over Oklab under both frameworks. For chroma and hue,
Oklch+ outperforms Oklab in predicting observer-rated discrimination, but Oklab
retains a marginal advantage in matching CAM16-UCS appearance proportionality.
It is noteworthy that CIEDE2000, not Oklch+, achieves the lowest STRESS on
chroma- and hue-dominant pairs in COMBVD (Table~\ref{tab:attribute}). The reversal
constitutes quantitative evidence that discrimination uniformity and appearance
uniformity can diverge, consistent with Section~\ref{sec:contrast}.
These attribute-level results should be interpreted as directional evidence;
the per-stratum sample sizes ($N = 122$--$142$) are not large enough to support
strong sub-group claims.

\subsection{Parameter Efficiency Relative to HELMLAB}

HELMLAB~\cite{Yildiz2026} achieves $\mathrm{STRESS} = 22.48$ on COMBVD with 72 free
parameters. On independent data from He et al.~\cite{He2022}, HELMLAB's STRESS (35.9)
exceeds CIEDE2000 (32.6), suggesting overfitting. Oklch+'s three-parameter structure
is analytically invertible, computationally trivial, and interpretable in terms of
established physiological mechanisms.

\subsection{The Naka-Rushton Transformation}

The Naka-Rushton function was originally proposed by Naka and Rushton~\cite{Naka1966}
to model the compressive response of retinal neurons to light intensity
(Eq.~\ref{eq:NR}). The same functional form appears in CIECAM02~\cite{CIE2004}
and CAM16~\cite{Li2017} as the compression applied to adapted cone responses.
Applying the NR function to Oklab chroma places Oklch+ in a lineage of models that
treat chroma sensitivity as a saturating process. The NR function is bounded in
$[0,1]$ for all $C \geq 0$, ensuring that the achromatic point is preserved
($C' = 0$ when $C = 0$)---a property the sigmoid function fails to satisfy
(Section~3.2).

\subsection{Relationship to CIEDE2000}

Despite operating through different mechanisms, Oklch+ and CIEDE2000 converge on
similar STRESS values (29.09 vs.\ 29.13). CIEDE2000 achieves this through
hue- and chroma-dependent weightings applied on top of CIELAB; Oklch+ achieves
comparable accuracy through two global, attribute-independent compressions applied to
Oklab. A key distinction is that Oklch+ defines a coordinate system in which
Euclidean distance approximates perceptual distance, enabling perceptually uniform
interpolation---a property that direct linear interpolation does not provide, as CIEDE2000 does not define a coordinate system.

% ─────────────────────────────────────────────────────────────────
\section{Conclusion}
% ─────────────────────────────────────────────────────────────────

We have proposed Oklch+, a three-parameter extension of Oklab comprising a power
transformation on the L-axis ($L' = L^{0.73}$, Eq.~\ref{eq:Lpow}) and a
Naka-Rushton compression on the C-axis
($C' = C^{0.87}/(C^{0.87} + 0.34^{0.87})$, Eq.~\ref{eq:NR}). Optimized against
COMBVD using the STRESS metric of Garc\'{\i}a et al.~\cite{Garcia2007}, Oklch+
achieves $\mathrm{STRESS} = 29.09$, closely matching CIEDE2000 (29.13;
difference $= 0.04$) with approximately 14 fewer parameters. Improvement over
Oklab (47.35) is confirmed across all six COMBVD sub-datasets.

Cross-validation on a held-out BFD-P D65 subset (2,028~pairs) yields
$\mathrm{STRESS} = 26.14$, substantially outperforming Oklab (51.45) and comparable
to CIEDE2000 (24.12). An additional experiment using CAM16-UCS appearance distances
as reference values reveals that Oklch+ outperforms Oklab on appearance-model ratio
uniformity across all pairs. Attribute-stratified analysis shows this advantage is
consistent for lightness but reverses for chroma and hue under appearance-model
evaluation, providing quantitative evidence that discrimination uniformity and
appearance uniformity can diverge.

The primary limitation is the chromaticity distribution of COMBVD (99.2\% of pairs
have $C < 0.20$). Under CAM16-UCS evaluation, Oklch+'s advantage is maintained
across wider gamuts including Display P3 and Rec.2020 (Table~\ref{tab:gamut}).
However, this constitutes theoretical validation based on an appearance model;
empirical validation with observer-rated discrimination data in the high-chroma
region remains future work. Additionally, CIEDE2000 retains a clear advantage on
chroma- and hue-dominant pairs (Table~\ref{tab:attribute}).

Because Oklch+ defines a coordinate system in which Euclidean distance approximates
perceptual distance, linear interpolation in the $(L', a', b')$ space produces
transitions with substantially improved perceptual uniformity relative to Oklab,
while also providing substantially better color difference prediction accuracy.

% ─────────────────────────────────────────────────────────────────
% Acknowledgments
% ─────────────────────────────────────────────────────────────────
\section*{Acknowledgments}

The author used Claude (Anthropic) as an AI writing assistant during the preparation
of this manuscript. The author takes full responsibility for the content.

% ─────────────────────────────────────────────────────────────────
% Data Availability
% ─────────────────────────────────────────────────────────────────
\section*{Data Availability Statement}

The COMBVD dataset used in this study is publicly available from
He et al.~\cite{He2022}. The Python scripts for parameter optimization and STRESS
evaluation are available from the corresponding author upon reasonable request.

% ─────────────────────────────────────────────────────────────────
% Conflict of Interest
% ─────────────────────────────────────────────────────────────────
\section*{Conflict of Interest Statement}

The author declares no conflicts of interest.

% ─────────────────────────────────────────────────────────────────
% References
% ─────────────────────────────────────────────────────────────────
\bibliographystyle{unsrtnat}
\bibliography{oklchplus_refs}

% ─────────────────────────────────────────────────────────────────
% Figure Captions
% ─────────────────────────────────────────────────────────────────
\clearpage
\section*{Figure Captions}

\textbf{Figure 1.}
Comparison of four candidate chroma transformation functions $C' = f(C)$.
The Sigmoid function does not satisfy $f(0) = 0$ at its optimal parameters
($a = 42$, $C_\mathrm{th} = 0.068$), disqualifying it as a chroma transformation
regardless of STRESS performance. The adopted Naka-Rushton (NR) function satisfies
$f(0) = 0$ for all $n, \sigma > 0$ and is bounded in $[0, 1]$.
All curves are shown at jointly optimized parameter values.

\textbf{Figure 2.}
STRESS by sub-dataset for Oklab, Oklch+, and CIEDE2000. Lower values indicate better
agreement with observer-rated color difference judgments. Dashed lines indicate ALL
COMBVD STRESS for each model. Sub-dataset sizes: BFD-P D65, $n = 2{,}028$;
BFD-P M, $n = 548$; BFD-P C, $n = 200$; Witt, $n = 418$; RIT-DuPont$^\dagger$,
$n = 312$; LEEDS, $n = 307$.
$^\dagger$RIT-DuPont $\Delta V$ values are constant across all pairs; scores for this
sub-dataset are not directly comparable to others.

\textbf{Figure 3.}
Distribution of Oklch chroma values $C$ in COMBVD (3,813~pairs;
$7{,}626$ individual color samples). A total of 99.2\% of pairs fall below
$C = 0.20$, indicating that the evaluation is largely confined to the low-to-mid
chroma range.

% ─────────────────────────────────────────────────────────────────
% Figures
% ─────────────────────────────────────────────────────────────────
\clearpage

\begin{figure}[ht]
  \centering
  \includegraphics[width=0.85\textwidth]{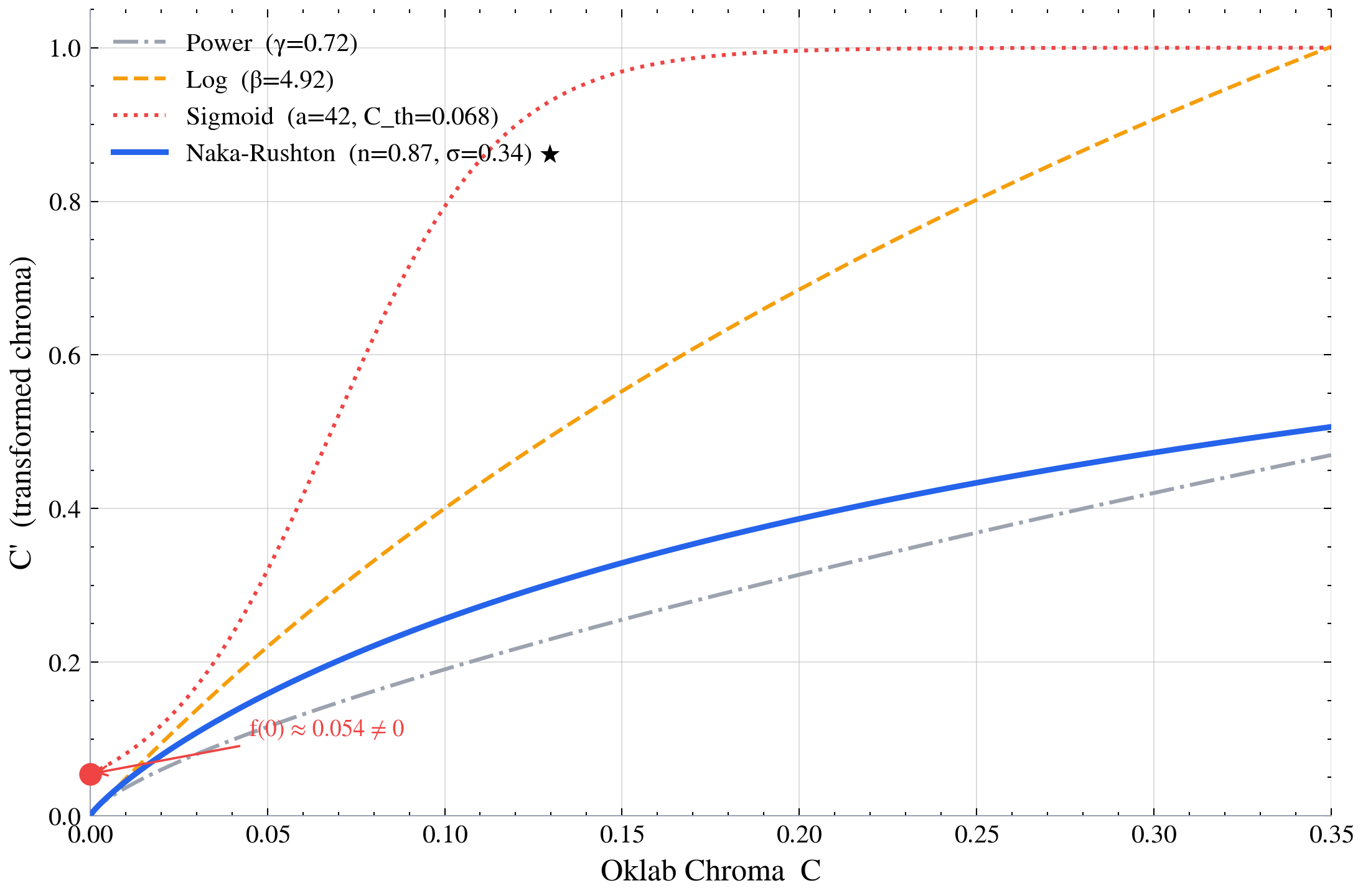}
  \caption{Comparison of four candidate chroma transformation functions $C' = f(C)$.
  The Sigmoid function does not satisfy $f(0) = 0$ at its optimal parameters,
  disqualifying it regardless of STRESS performance. The adopted Naka-Rushton (NR)
  function satisfies $f(0) = 0$ for all $n, \sigma > 0$ and is bounded in $[0,1]$.}
  \label{fig:c_transform}
\end{figure}

\begin{figure}[ht]
  \centering
  \includegraphics[width=0.92\textwidth]{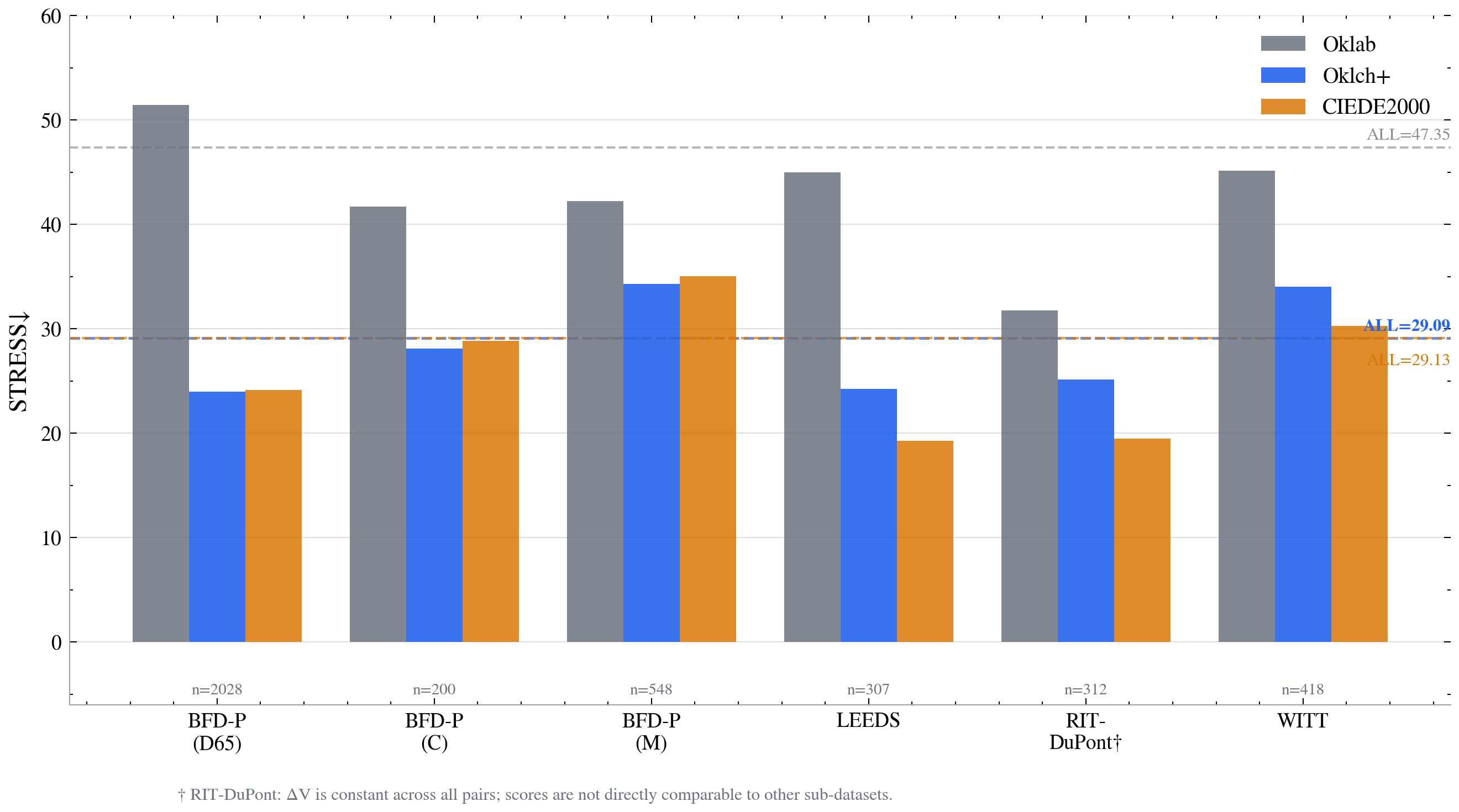}
  \caption{STRESS by sub-dataset for Oklab, Oklch+, and CIEDE2000.
  Dashed lines indicate ALL COMBVD STRESS for each model.
  $^\dagger$RIT-DuPont $\Delta V$ values are constant; scores are not directly
  comparable to other sub-datasets.}
  \label{fig:substress}
\end{figure}

\begin{figure}[ht]
  \centering
  \includegraphics[width=0.80\textwidth]{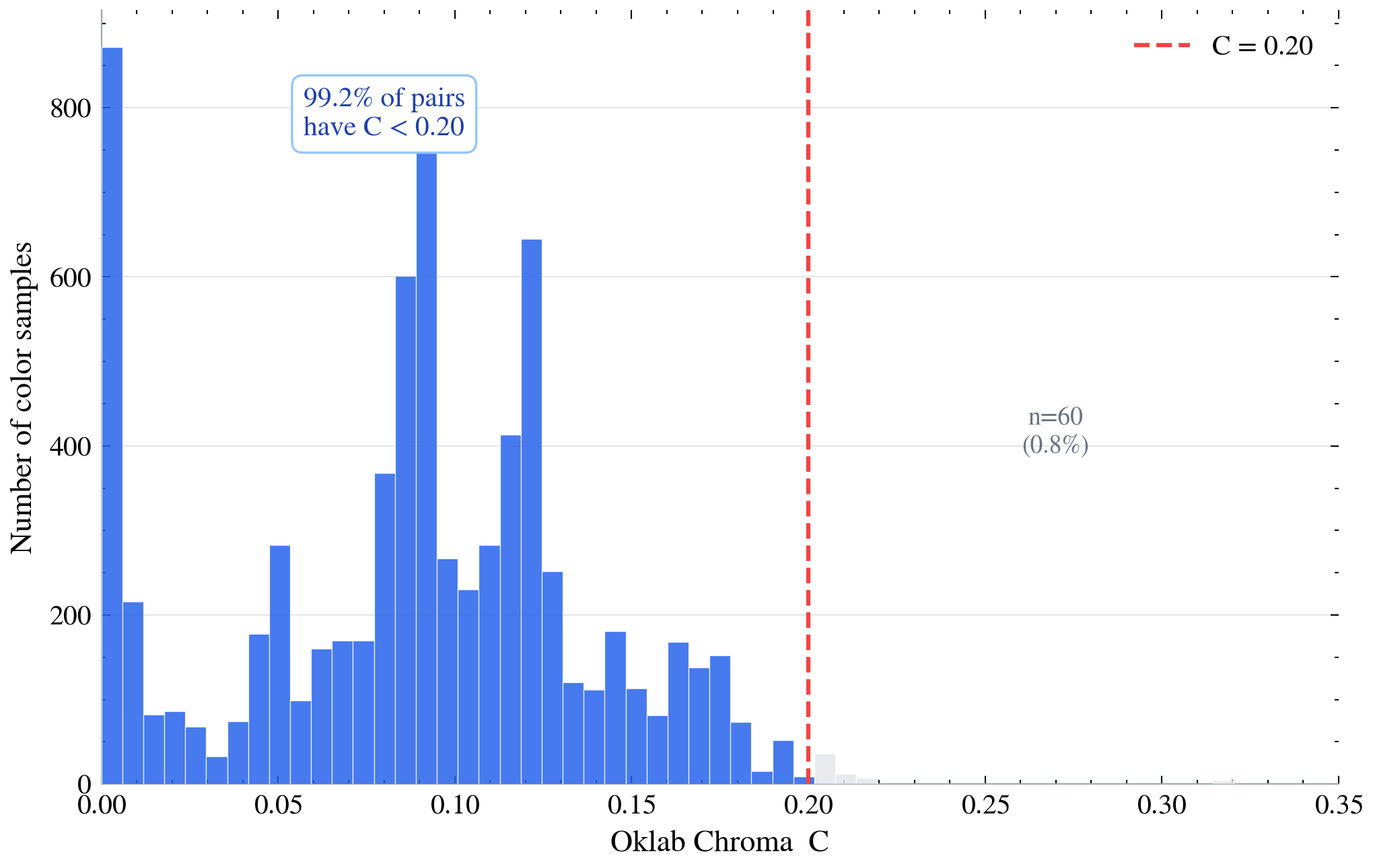}
  \caption{Distribution of Oklch chroma values $C$ in COMBVD (3,813~pairs).
  A total of 99.2\% of pairs fall below $C = 0.20$, indicating evaluation is
  largely confined to the low-to-mid chroma range.}
  \label{fig:chroma_dist}
\end{figure}

\end{document}